\documentclass[runningheads]{llncs}

\usepackage[utf8]{inputenc}
\usepackage[T1]{fontenc}

\let\subparagraph\relax
\usepackage{titlesec}

\usepackage{amsmath}
\usepackage{amsfonts}
\usepackage{algpseudocode}
\usepackage{empheq}

\usepackage{graphicx}
\usepackage[dvipsnames,table]{xcolor}  

\usepackage{booktabs}
\usepackage{colortbl}
\usepackage{multirow}
\usepackage{multicol}

\usepackage{tocloft}
\usepackage{mdwlist}
\usepackage{fancyhdr}
\usepackage{moresize}
\usepackage{placeins}

\usepackage{nomencl}

\makenomenclature

\usepackage{lipsum}

\usepackage{hyperref}
\usepackage{xurl}

\usepackage{fancyhdr}

\fancypagestyle{firstpage}{
  \fancyfoot[C]{\scriptsize14th International Conference on
Security, Privacy and Applied Cryptographic Engineering 2024}

  \fancyhead{}
}
\begin{document}
\title{Privacy-Preserving Graph-Based Machine Learning with Fully Homomorphic Encryption for Collaborative Anti-Money Laundering}
\titlerunning{Privacy-Preserving Graph-Based ML with FHE for Collaborative AML}
%
\author{Fabrianne Effendi\inst{1}\orcidID{0009-0002-8595-4323} \and
Anupam Chattopadhyay\inst{1}\orcidID{0000-0002-8818-6983} }
\authorrunning{F. Effendi and A. Chattopadhyay}
%
\institute{Nanyang Technological University, 50 Nanyang Ave, 639798, Singapore \\
\email{fabr0001@e.ntu.edu.sg, anupam@ntu.edu.sg}}
\maketitle              
\thispagestyle{firstpage}

\begin{abstract}
Combating money laundering has become increasingly complex with the rise of cybercrime and digitalization of financial transactions. Graph-based machine learning techniques have emerged as promising tools for Anti-Money Laundering (AML) detection, capturing intricate relationships within money laundering networks. However, the effectiveness of AML solutions is hindered by data silos within financial institutions, limiting collaboration and overall efficacy. This research presents a novel privacy-preserving approach for collaborative AML machine learning, facilitating secure data sharing across institutions and borders while preserving privacy and regulatory compliance. Leveraging Fully Homomorphic Encryption (FHE), computations are directly performed on encrypted data, ensuring the confidentiality of financial data. Notably, FHE over the Torus (TFHE) was integrated with graph-based machine learning using Zama Concrete ML. The research contributes two key privacy-preserving pipelines. First, the development of a privacy-preserving Graph Neural Network (GNN) pipeline was explored. Optimization techniques like quantization and pruning were used to render the GNN FHE-compatible. Second, a privacy-preserving graph-based XGBoost pipeline leveraging Graph Feature Preprocessor (GFP) was successfully developed. Experiments demonstrated strong predictive performance, with the XGBoost model consistently achieving over 99\% accuracy, F1-score, precision, and recall on the balanced AML dataset in both unencrypted and FHE-encrypted inference settings. On the imbalanced dataset, the incorporation of graph-based features improved the F1-score by 8\%. The research highlights the need to balance the trade-off between privacy and computational efficiency.

\end{abstract}
%
%
%
\section{Introduction}
\subsection{Background and Motivation}
\subsubsection{The Money Laundering Problem.} 
Money laundering is the process of concealing the origins of illegally obtained funds to make them appear legitimate. It poses severe consequences, fostering corruption, organized crime, terrorism, and environmental offenses. Beyond discouraging foreign investments and distorting international capital flows, money laundering poses a substantial threat to the stability of the financial system and the broader economy. The United Nations estimates that 2-5\% of global GDP is laundered annually, with only 1\% of these funds seized \cite{un2023}. For example, in 2023, Singapore uncovered a money laundering case involving over S\$3 billion (US\$2.21 billion), making it one of the largest cases discovered globally \cite{cheah_2024}. 

\subsubsection{Graph-Based Machine Learning in Anti-Money Laundering (AML).} 
 Money laundering activities inherently involve network structures, where their web of transactions can be modelled by graph-like patterns such as fan-out, fan-in, gather-scatter and so on \cite{altman2023realistic}. Machine learning have gained considerable interest in uncovering hidden patterns in transaction data \cite{mckinsey_fight_2022}. Particularly, graph-based machine learning \cite{kurshan2021graph} including graph neural networks (GNNs) show great promise in AML detection (section \ref{sec:gnn_lit_review}). They can effectively capture complex relationships between entities such as individuals, businesses and accounts, identifying money-laundering patterns, learning anomalies, and unveiling hidden connections that are often missed by traditional detection methods.

\subsubsection{The Challenge of AML Silos.}
Despite the advancements in AML solutions, AML transaction monitoring and detection are often conducted in silos within each financial institution \cite{pwc_breaking_nodate}, hindering the effectiveness of AML solutions. Adherence to data privacy regulations, such as the EU's General Data Protection Regulation (GDPR) \cite{wass_banks_2020}, poses legal challenges to financial institutions in maintaining the privacy of transaction data when data sharing. It is not possible for financial institutions to directly share meta-information about transaction accounts or owners without explicit consent. Consequently, this prevents the comprehensive modeling of transaction networks and limits the ability of financial institutions to track transactions and customer relationships spanning across multiple institutions and borders, which is often exploited by criminals orchestrating intricate schemes across multiple institutions. 

\subsubsection{Collaborative AML.} 
To effectively combat money laundering, a collaborative approach is required. Emerging solutions involve leveraging a Trusted Third Party (TTP) to aggregate and analyze data, while preserving privacy. For example, the Monetary Authority of Singapore (MAS) developed COSMIC in Singapore \cite{mas_cosmic_2023}, a platform for collaborative sharing of Money Laundering/Terrorism Financing (ML/TF) information among banks using rule-based transaction monitoring \cite{mas_consultation_2023}. However, many institutions still lack legally-compliant infrastructure for secure data sharing. According to a FATF report, regulatory challenges and data privacy concerns were the two most frequently cited challenges in developing and implementing new AML technologies, with nearly 70\% and 60\% of respondents highlighting these issues respectively \cite{fatf2021}.

\subsubsection{Privacy-Preserving Technologies.} 
Privacy-preserving technologies, such as Homomorphic Encryption (HE), offer a promising solution. HE allows computations to be performed directly on encrypted data without decryption, ensuring data confidentiality. The increasing adoption and community support of HE, such as Fully Homomorphic Encryption over the Torus (TFHE) (section \ref{sec:fhe-literature}), motivates research in applying it to AML to enable secure collaborative efforts.

\subsection{Objective and Contributions} 
This paper presents a novel privacy-preserving AML approach using TFHE to enable secure data sharing and collaboration between financial institutions while protecting data privacy. 
To the best of my knowledge, while there are recent research works that have conducted separate explorations of graph-based machine learning for AML (section \ref{sec:gnn_related_works}), Gradient-Boosting Tree (GBT) for AML (section \ref{sec:gradient_boosting_related_works}), privacy-preserving technologies (PPTs) for collaborative AML and privacy-preserving machine learning respectively (section \ref{sec:PETs}), none have combined these areas of study to explore the use of GNN or GBT with FHE in AML, all the less so with the TFHE scheme. Hence, this paper makes the following contributions:
\begin{enumerate}
    \item A proposed solution architecture for collaborative privacy-preserving machine learning for AML using FHE. This is detailed in section \ref{sec:aml_fhe_architecture}.
    \item Exploration of the feasibility of developing a privacy-preserving GNN pipeline integrating TFHE using Concrete ML \cite{zama_what_2024} for money laundering detection. This exploration is detailed in section \ref{sec:gnn_pipeline}.
    \item Development of a privacy-preserving graph-based XGBoost pipeline leveraging the Graph Feature Preprocessor (GFP) \cite{blanuša2024graph} and integrating TFHE using Concrete ML \cite{zama_what_2024} to detect money laundering. This is detailed in section \ref{sec:xgb_gfp_pipeline}. 
    \item A set of experiments with incrementally GFP-enriched graph features using XGBoost, comparing the model performance for both unencrypted and TFHE-encrypted inference, and against a basic XGBoost baseline. The results evaluated the trade-offs of building a privacy-preserving pipeline between privacy and model performance. This is detailed in section \ref{chapt:experiments}. 
    
\end{enumerate}

We have also released the project code repository on GitHub\footnote{https://github.com/fabecode/GraphML-FHE}.

\section{Related Work}\label{sec:lit_review}
\subsection{Graph-Based Machine Learning for AML}\label{sec:gnn_lit_review}\label{sec:gnn_related_works}
The application of Graph Neural Networks (GNNs) to the domain of Anti-Money Laundering (AML) has been an active area of research. Weber et al. conducted early experiments on synthetic AML datasets, finding that variants like Fast Graph Convolutional Networks (FastGCN) offered faster computation \cite{weber2018scalable}. They later explored GNNs and traditional machine learning methods on the real-world Elliptic dataset, with GCN outperforming simpler models but being surpassed by random forest \cite{weber2019antimoney}. Recognizing the importance of temporal information, Alarab et al. \cite{alarab_graph-based_2023} and Pareja et al. \cite{pareja2019evolvegcn} developed novel GNN architectures that captured evolving financial transaction patterns over time on the Elliptic dataset. Cardoso et al. \cite{cardoso2022laundrograph} introduced LaundroGraph, a self-supervised graph representation learning approach tailored for AML, showcasing the potential of graph-based techniques in extracting meaningful representations from financial interaction networks. Building on these foundational works, Johannessen et al. explored heterogeneous graph neural networks for real-world AML scenarios at DNB Bank \cite{johannessen2023finding}, while Altman et al. \cite{altman2023realistic} and Egressy et al. \cite{egressy2024provably} made significant contributions in developing and evaluating GNN models on the comprehensive synthetic AMLworld dataset. This body of research consistently highlights the effectiveness of GNNs in capturing the complex, graph-structured nature of money laundering activities, paving the way for continued advancements in this field.

\subsection{Gradient-Boosted Trees (GBT) for AML}\label{sec:gbt_lit_review}\label{sec:gradient_boosting_related_works}
The application of gradient-boosted tree (GBT) models, such as XGBoost and LightGBM, has also gained significant traction, with several studies including Tertychnyi et al. \cite{tertychnyi2020aml}, Jullum et al. \cite{julum2020aml} and Vassallo et al. \cite{vassallo2021aml}, demonstrating the effectiveness of these ensemble learning techniques in AML detection. More recently, researchers have explored integrating graph-based feature extraction techniques with GBT models to further enhance their performance in AML tasks. Eddin et al. \cite{eddin2022antimoney} showcased the benefits of incorporating graph-based features, such as degree and GuiltyWalker, into LightGBM models. Building on this, Altman et al. \cite{altman2023realistic} and Blanuša et al. \cite{blanuša2024graph} conducted comprehensive comparisons of GNN baselines and GBT models enhanced with their Graph Feature Preprocessor (GFP) on the AMLworld dataset, finding that the GBT-GFP combination, particularly XGBoost-GFP, outperformed the GNN approaches. These advancements highlight the potential of leveraging the synergies between graph-based representations and gradient-boosted tree models for effective AML detection.

\subsection{Privacy-Preserving Technologies (PETs)}\label{sec:PETs} 
\subsubsection{PETs in Financial Crimes.}\label{sec:PETs_finance}
Privacy-Preserving Technologies (PETs) have emerged as a promising approach to protect sensitive information and enable secure data computation and analysis while preserving privacy in various domains, including finance applications \cite{cryptoeprint:2023/122}. The integration of PETs with AML efforts is a relatively new and active area of research.

One notable industry-led proof-of-concept is Project Aurora by the Bank for International Settlements \cite{bis_innovation_hub_project_2023}, which explored the use of different PET combinations, along with machine learning and network analysis, to detect money laundering across siloed, national, and cross-border payment data. Their findings suggest that a centralized cross-border approach leveraging HE and Local Differential Privacy (LDP) demonstrates the best performance in AML. Meanwhile, Mastercard showcased the potential of FHE in facilitating secure cross-border sharing of financial crime intelligence among Singapore, the United States, India, and the United Kingdom \cite{mastercard_imda_pet_2023}.

Recently, research-focused advancements have also emerged in this space. Zhang et al. \cite{zhang2023privacypreserving} proposed a framework that utilizes hybrid Federated Learning (FL) to enable collaborative financial crime detection among multiple institutions, while Egmond et al. \cite{cryptoeprint:2024/065} developed a secure risk propagation algorithm using secure multi-party computation (MPC) and additive HE for confidential risk score updates across a collaborative inter-bank network. 

\subsubsection{PETs for Privacy-Preserving Machine Learning.}\label{sec:PETs_ML}
Numerous studies have also explored the integration of FHE with machine learning models, primarily focusing on Convolutional Neural Networks (CNNs), including CryptoNets by Dowlin et al. \cite{pmlr-v48-gilad-bachrach16}, Homomorphic CNN (HCNN) by Badawi et al. \cite{cryptoeprint:2018/1056}, Low-Latency CryptoNets (LoLa) by Brutzkus et al. \cite{brutzkus2019low}, and ResNet-20 with RNS-CKKS FHE and bootstrapping by Lee et al. \cite{lee2021privacypreserving}. However, these approaches often prove ineffective for Graph Neural Networks (GNNs) due to the differences in computational patterns between the two. While FHE-based CNN inference often focuses on optimizing 2D convolutions, GNNs like Graph Convolutional Networks (GCNs) introduce different computational patterns. Specifically, GCNs rely heavily on consecutive matrix multiplications for feature and node aggregation, creating a bottleneck that is not present in CNN layers, where multi-channel 2D convolutions dominate the computations \cite{ran2023penguin}.

Notably, in the area of privacy-preserving GNNs, Ran et al. \cite{ran2022cryptogcn} introduced CryptoGCN, a HE-based framework for Graph Convolutional Network (GCN) inference, leveraging the sparsity of matrix operations to minimize encrypted computational overhead. These advancements in PET-based approaches underscore the growing importance and potential of privacy-preserving techniques in the fight against financial crimes, setting the stage for the proposed privacy-preserving AML solution in this research.

\section{Homomorphic Encryption (HE)}\label{sec:he-literature}
\subsection{HE Overview}
Homomorphic Encryption (HE) is a cryptographic technique that allows computations to be directly performed  on encrypted data without the need for decryption. Unlike traditional encryption, which necessitates the decryption of data before meaningful computation, HE empowers secure computation while maintaining the confidentiality of data throughout the computation process. 

\subsection{Fully Homomorphic Encryption (FHE)}\label{sec:fhe-literature}
Fully Homomorphic Encryption (FHE) was imagined by Rivest et al. \cite{rivest1978data} in 1978, and the first scheme was developed by Craig Gentry \cite{10.5555/1834954} in 2009. FHE supports an unlimited number of computations on encrypted data without revealing the underlying messages. Mathematically, if a user has an arbitrary function \(f\) and aims to derive \(f\left(m_1, \ldots, m_n \right)\) for some inputs \(m_1, \ldots, m_n\), the plaintexts \(p_1, \ldots, p_n\) are first encrypted using the public key \(pk\), i.e.: 
\begin{equation}
Encrypt\left(pk, p_1, \ldots, p_n\right) = c_1, \ldots, c_n
\vspace{-3pt}
\end{equation}

Computations are then performed directly on the encrypted ciphertexts \(c_1, \ldots, c_n\). Decryption of the output using secret key \(sk\) gives the result \(f\left(m_1, \ldots, m_n \right)\), i.e.:
\begin{equation}
Decrypt\left(sk, Eval\left(pk,f,c_1, \ldots, c_n \right)\right) = f\left(m_1, \ldots, m_n \right)
\vspace{-6pt}
\end{equation}

\subsubsection{Machine Learning using FHE.}\label{sec:fhe_with_ml_lit_review}
In the context of machine learning, a party can encrypt input data using the public key, and the model will be able to process the data without seeing the original data. The final result is also encrypted, and the data can only be decrypted by the client who has the private key. This ensures confidentiality of the data from the machine learning model.

\begin{figure}[h]
\vspace{-10pt}
  \centering
  \setlength{\fboxsep}{1pt} 
  \setlength{\fboxrule}{1pt} 
  \fbox{\includegraphics[width=0.95\textwidth]{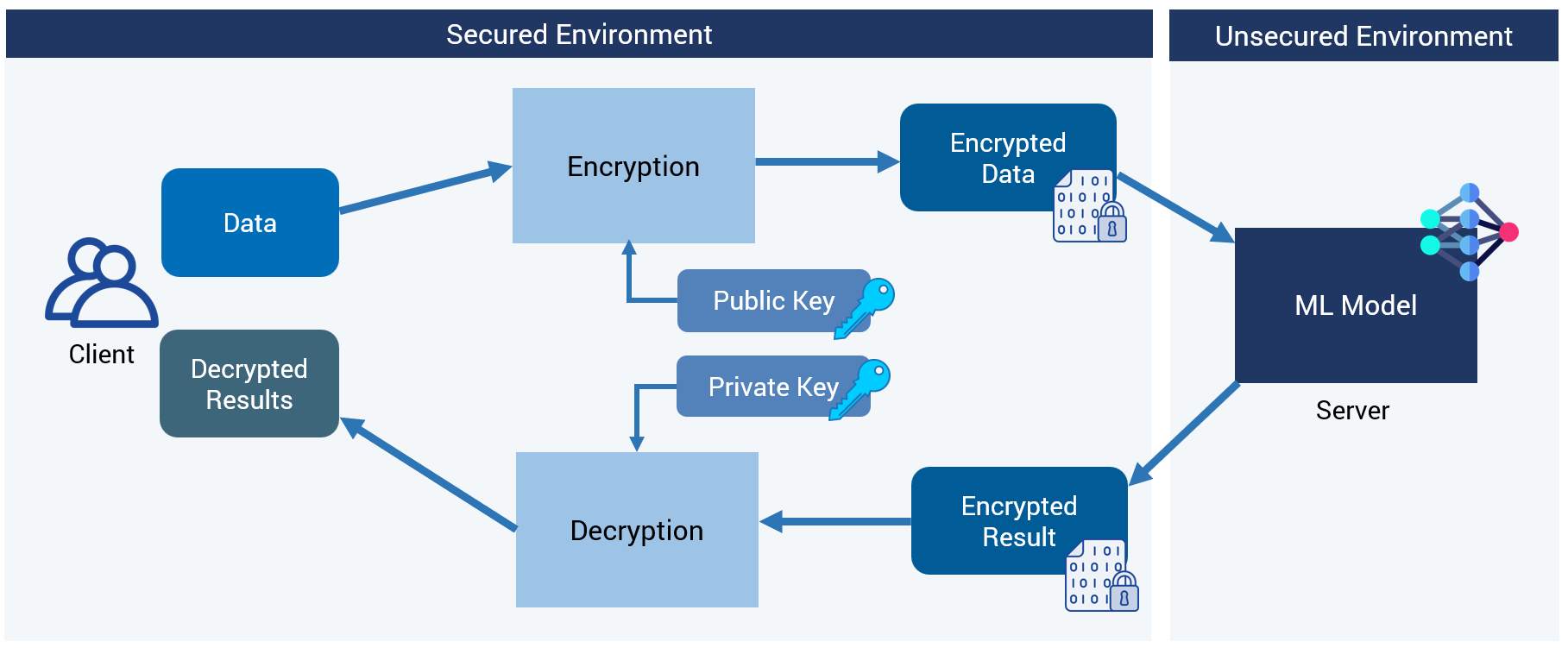}}
  \vspace{-6pt}
  \caption{Fully Homomorphic Encryption Process in Machine Learning}
  \label{fig:fhe_process}
\end{figure}
\vspace{-0.7cm}

\subsubsection{Fully Homomorphic Encryption over the Torus (TFHE).}
TFHE is a specialized FHE scheme designed by Chillotti et al. \cite{cryptoeprint:2018/421} for efficient computation of Boolean circuits on encrypted data. It enables very fast gate bootstrapping as well as circuit bootstrapping and operations over Boolean gates, reducing bootstrapping time to 13\(ms\).

\subsubsection{TFHE-Concrete.}TFHE-Concrete, an extended version of TFHE in Chillotti et al. \cite{chillotti2020concrete}, enhances the versatility and performance of TFHE in practical applications. It pushes the frontiers of bootstrapping by being the first to implement programmable bootstrapping (PBS), allowing for the simultaneous evaluation of arbitrary univariate function during bootstrapping. This is achieved by replacing the plaintext bits with a function of them in a lookup table. Presently, this is the most powerful technique for efficiently evaluating homomorphic non-linear functions, including activation functions in deep neural networks \cite{zama_he_101_2021}. By combining the benefits of fast bootstrapping with programmable bootstrapping functionality, TFHE-Concrete enables secure and efficient computations on encrypted data, making it a valuable tool for privacy-preserving data analysis. 

\subsection{Zama's Concrete Framework}\label{sec_concrete}
Zama's Concrete framework is an open-source tool that empowers developers to integrate HE into their applications without the need for in-depth cryptography knowledge \cite{chillotti2020concrete}. TFHE-Concrete, as an integral part of the Concrete framework, provides comprehensive support across various categories, including leveled operations, bootstrapped operations, and PBS. It supports the approximate or exact evaluation of arbitrary functions, and supports both Boolean and integer operations, making it a versatile and integrated FHE solution.

\begin{table}[ht]
    \vspace{-6pt}
  \centering
  \setlength{\tabcolsep}{2pt}
  \caption{Comparison of TFHE-Concrete with Other FHE Schemes \cite{zama_concrete_framework_2022}}. 
  \begin{tabular}{lcccccccc}
    \toprule
    FHE Schemes & \multicolumn{3}{c}{Operations} & \multicolumn{2}{c}{Non-linear} & \multicolumn{3}{c}{Data Types}\\
    \cmidrule(lr){2-4} \cmidrule(lr){5-6} \cmidrule(lr){7-9}
    & Level & Bootstrapped & PBS & Exact & Approx & Bool & Int & Real \\
    \midrule
    \textbf{BGV} & $\checkmark$ & $\times$ & $\times$ & $\times$ & $\checkmark$ & $\times$ & $\checkmark$ & $\times$ \\
    \textbf{BFV} & $\checkmark$ & $\times$ & $\times$ & $\times$ & $\checkmark$ & $\times$ & $\checkmark$ & $\times$ \\
    \textbf{CKKS} & $\checkmark$ & $\times$ & $\times$ & $\times$ & $\checkmark$ & $\times$ & $\times$ & $\checkmark$\\
    \textbf{TFHE-Lib} & $\times$ & $\checkmark$ & $\times$ & $\checkmark$ & $\times$ & $\checkmark$ & $\times$ & $\times$\\
    \textbf{TFHE-Concrete} & $\checkmark$ & $\checkmark$ & $\checkmark$ & $\checkmark$ & $\checkmark$ & $\checkmark$ & $\checkmark$ & $\times$\\
    \bottomrule
  \end{tabular}
  \label{fhe_scheme_comparison}
\end{table}
\vspace{-1cm}

\subsubsection{Concrete ML.}
Concrete ML is a privacy-preserving machine learning Python toolkit built on top of the Concrete framework \cite{zama_what_2024}. Building on Concrete's TFHE and PBS implementation, it empowers data scientists to develop privacy-preserving models on encrypted data using the TFHE framework for secure machine learning inference.

The toolkit incorporates ready-to-use FHE-friendly models with an interface equivalent to scikit-learn. Additionally, it also provides support for customs models, including deep neural networks built with PyTorch or Keras/Tensorflow. For custom models, it is necessary to implement quantization before compiling to FHE, utilizing third-party libraries like Brevitas for PyTorch. The model is subsequently converted and imported into Concrete ML through the Open Neural Network Exchange (ONNX), an open-source standard facilitating interoperability across various deep learning frameworks and hardware platforms.

\subsection{FHE Implementation using Concrete ML}
To implement FHE in our pipelines, Concrete ML was leveraged as the FHE implementation library. Concrete ML was chosen for its suitability in realizing privacy-preserving FHE-based machine learning models and its user-friendly interface for data scientists to develop privacy-preserving models on encrypted data. Training was first done on unencrypted data, producing a model that was then converted to an FHE equivalent that can perform encrypted inference.

In the project, Concrete ML is used in 2 forms:
\begin{enumerate}
    \item For XGBoost, Concrete ML's built-in scikit-learn-like interface is leveraged to build an FHE-compatible XGBoost that performs encrypted inference.
    \item For GNN, a custom model architecture is created using PyTorch Geometric, while leveraging Concrete ML for compilation, encryption, and inference.
\end{enumerate}

\subsection{Collaborative AML with FHE Solution Architecture}\label{sec:aml_fhe_architecture}

Below is a high-level overview of the project's collaborative AML with FHE solution architecture.

\begin{figure}[ht]
\vspace{-6pt}
  \centering
  \setlength{\fboxsep}{1pt} 
  \setlength{\fboxrule}{1pt} 
  \fbox{\includegraphics[width=1.0\textwidth]{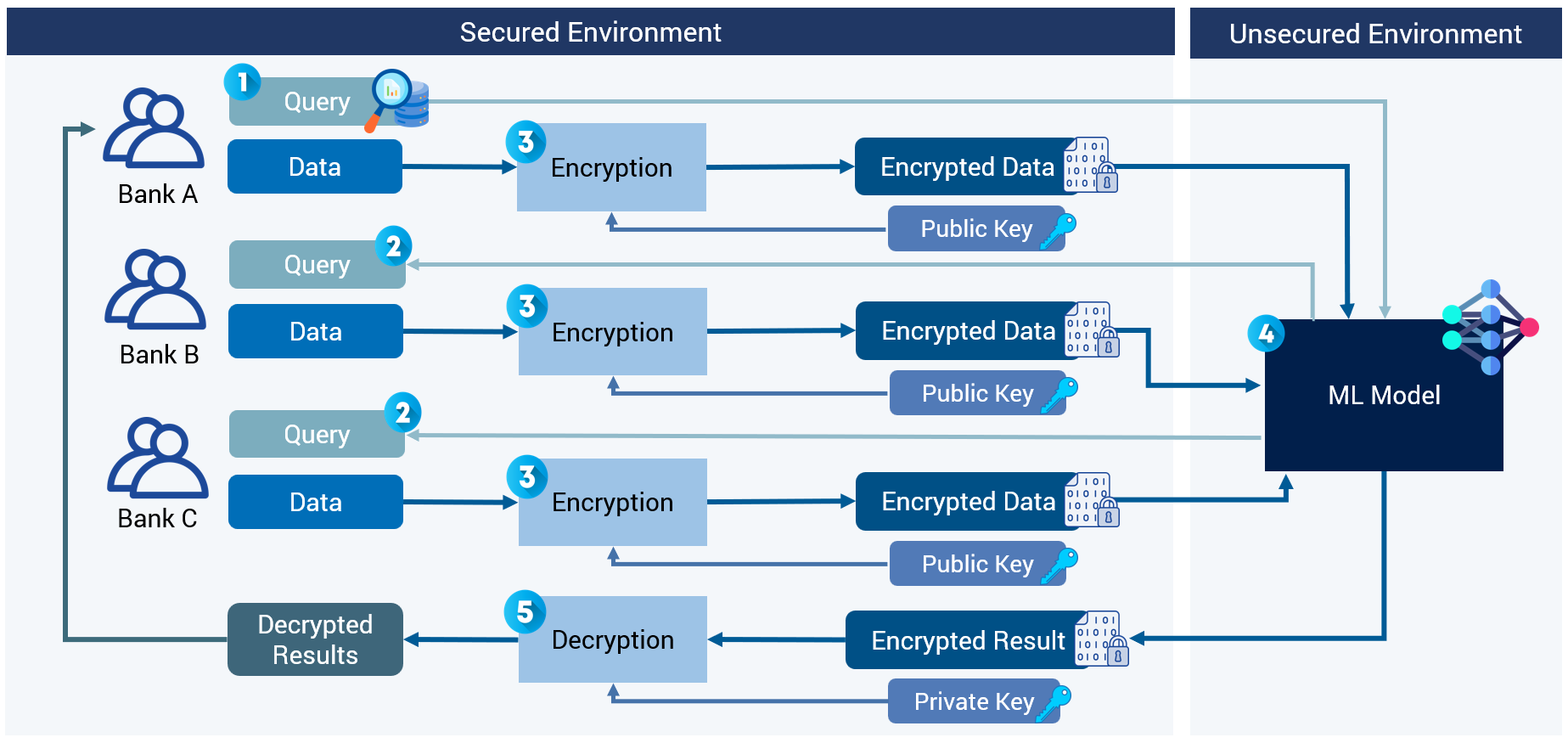}}
  \vspace{-0.5cm}
  \caption{Collaborative FHE Solution Architecture}
  \label{fig:FHE-MPC}
\end{figure}

\begin{enumerate}
    \item \textbf{Client Initialization:} A financial institution (FI) initiates the AML computation process with a query.

    \item \textbf{Query Forwarding:} The centralised server forwards the query to participating FIs.

    \item \textbf{Input Encryption:} Participating FIs, including the inquiry FI, encrypt relevant data using FHE with the inquiry FI’s public key. This key is securely shared once, allowing for multiple transactions to be encrypted without repeated exchanges.
    
    \item \textbf{Secure Computation:} The consolidated encrypted data from multiple parties undergoes machine learning inference on the server, producing results in encrypted form while maintaining input confidentiality.

    \item \textbf{Result Decryption:} The final encrypted results are decrypted by the inquiry FI using its private key, revealing only the output of the computation while keeping the individual inputs from participating FIs confidential.

\end{enumerate}

The above process aligns with confidentiality regulations, such as those in Singapore, where the Banking Act prohibits the disclosure of Customer Information (CI), or the existence of a non-public relationship between the customer and the bank \cite{mastercard_imda_pet_2023}.

\section{TFHE-Compatible GNN Pipeline}\label{sec:gnn_pipeline}
\begin{figure}[h]
\vspace{-0.5cm}
  \centering
  \setlength{\fboxsep}{1pt} 
  \setlength{\fboxrule}{1pt}
  \fbox{\includegraphics[width=0.98\textwidth]{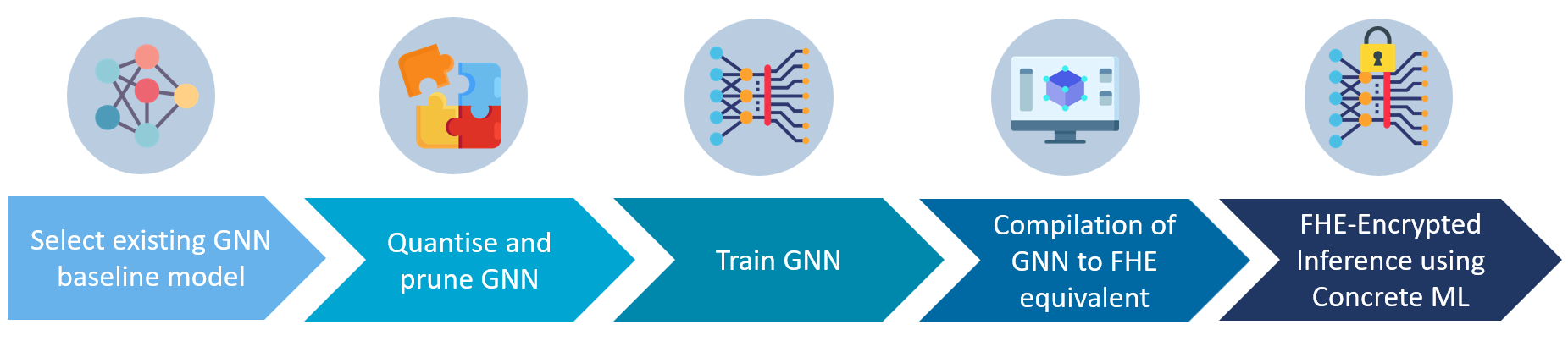}}
  \vspace{-6pt}
  \caption{Privacy-Preserving GNN Pipeline}
  \label{fig:fhe-gnn-pipeline}
  \vspace{-12pt}
\end{figure}

The GNN pipeline focused on exploring the feasibility of making a Graph Neural Network (GNN) compatible with Fully Homomorphic Encryption (FHE), a challenge that has not been extensively addressed due to the inherent complexity of GNNs. This aimed to address the lack of prior work in integrating GNNs with FHE, which is crucial for enabling privacy-preserving collaborative machine learning in domains such as anti-money laundering. 

The pipeline involved: selecting a GNN baseline model, quantizing and pruning it for FHE compatibility, training on Concrete ML, compiling the GNN to its FHE equivalent, and finally performing FHE inference.

\subsection{Baseline GNN model}
The Graph Isomorphism Network (GIN) was selected as the baseline GNN model. GIN is a message passing GNN that uses an iterative aggregation mechanism inspired by the Weisfeiler-Lehman graph isomorphism test \cite{xu2019powerful}. The baseline GIN model is implemented using PyTorch Geometric and adapted from previous AML studies \cite{altman2023realistic,egressy2024provably,altman_multi-gnn_2023}.

\subsection{Quantization}
Quantization makes models TFHE-compatible by transforming floating-points to integer representation. It is the process of constraining an input from a continuous or otherwise large set of values to a discrete set. While primarily used for enhancing the efficiency and compression of neural networks, quantization also helps address specific limitations associated with with FHE \cite{zama_quantization_2022}. Firstly, quantization effectively substitutes floating-point value multiplications with integer multiplications, a feasible operation in FHE. Moreover, quantization enables the reduction of values to small integers, offering a strategy to navigate the challenges posed by limited precision in programmable bootstrapping.

\subsubsection{Quantization Targets.}
To refine the efficiency and security of Graph Neural Networks (GNNs), the following quantization targets strategies are employed to facilitate the quantization process. Each strategy focuses on specific aspects of the GNN architecture, encompassing both node and edge features, as well as the quantization of weights and activations within GNN layers.
\begin{enumerate}
\item \textbf{Node Feature Quantization:}
Node feature quantization reduces the precision of input node features. To achieve this, the chosen methodology involves the application of quantization to node features, effectively transforming their representation into lower-precision numerical values. In practice, this quantization is implemented through the adoption of quantized linear layers, such as QuantLinear in Brevitas, strategically incorporated into both input and hidden layers.

\item \textbf{Edge Feature Quantization:}
Quantizing edge features becomes imperative, particularly in scenarios involving edge convolutional layers. The methodology extends the quantization process to edge features, dependent on their relevance in the GNN architecture. Implementation-wise, the integration of quantized linear layers (QuantLinear) or alternative quantized operations is recommended to effectively process edge features.

\item \textbf{Weights and Activations Quantization:}
In the realm of GNN layers, the overarching objective is to quantize both weights and activations. This is achieved through substituting the traditional linear layers with quantized linear layers (QuantLinear), and activation function with quantized activation functions (QuantReLU) within the GNN model. 
\end{enumerate}

Overall, the quantization strategies encompass a reduction in weight bit width for layer configurations, and a decrease in accumulator bit width for activation. These quantization techniques contribute significantly to the optimization and security enhancement of Graph Neural Networks.

\subsubsection{Quantization-Aware Training (QAT).} Quantization-Aware Training (QAT) was employed to ensure the GNN model's compatibility with the reduced precision requirements of Fully Homomorphic Encryption (FHE). In QAT, the neural network undergoes training with an awareness of the quantization process, enabling the model to learn and adapt to the intricacies associated with lower bit-width weights and activations \cite{deepnnwithtfhe2023,tailor2021degreequant}. QAT was implemented using Brevitas for Pytorch. Each PyTorch layer was mapped to its quantized version in Brevitas, and the corresponding weight and bit width were configured accordingly.

\subsection{Pruning}
Pruning was employed to optimize the GNN model size and computational complexity, mitigating the risk of accumulator overflow during Fully Homomorphic Encryption (FHE) computations in Concrete ML \cite{deepnnwithtfhe2023}. Pruning involves setting certain weights in the neural network to zero, reducing the model's size and computational demands.

Some key benefits of pruning in the context of FHE include: controlling the number of active neurons to reduce computational complexity, managing the accumulator bit width to ensure compatibility with FHE's limited precision, and enhancing resource efficiency by reducing storage and computation requirements.

In the GNN implementation, edge pruning was performed on less informative edges to streamline graph connectivity while ensuring the integrity of the financial transactions graph and preserving critical components.

\subsection{Compilation of GNN model to FHE Equivalent}
Final conversion of the GNN model is performed by Concrete ML, which uses the Concrete Compiler to translate the Multi-Level Intermediate Representation (MLIR) representation of the model into an FHE program, generating machine code that executes the model on encrypted data. 

The conversion process starts with importing the ONNX model, followed by a sequence of transformations: initially into a NumpyModule, then into a QuantizedModule, and ultimately into an FHE circuit \cite{zama_compilation_2024}.  

While efforts were made to develop a privacy-preserving GNN pipeline using Concrete ML, challenges in integrating PyTorch Geometric with the Concrete ML framework limited the successful completion of this component. The absence of support for the ScatterElements ONNX operator in Concrete ML posed a significant challenge during the final compilation of Graph Neural Networks (GNNs). While a custom implementation was developed to fill this gap, ongoing challenges remain in the conversion of NumpyModule to Quantized module. Nevertheless, the progress made in implementing techniques to render the GNN TFHE-compatible provides a good foundation for future research.
\section{TFHE-Compatible XGBoost with GFP Pipeline}\label{sec:xgb_gfp_pipeline}
\begin{figure}[h]
\vspace{-0.5cm}
  \centering
  \setlength{\fboxsep}{1pt} 
  \setlength{\fboxrule}{1pt}
  \fbox{\includegraphics[width=0.98\textwidth]{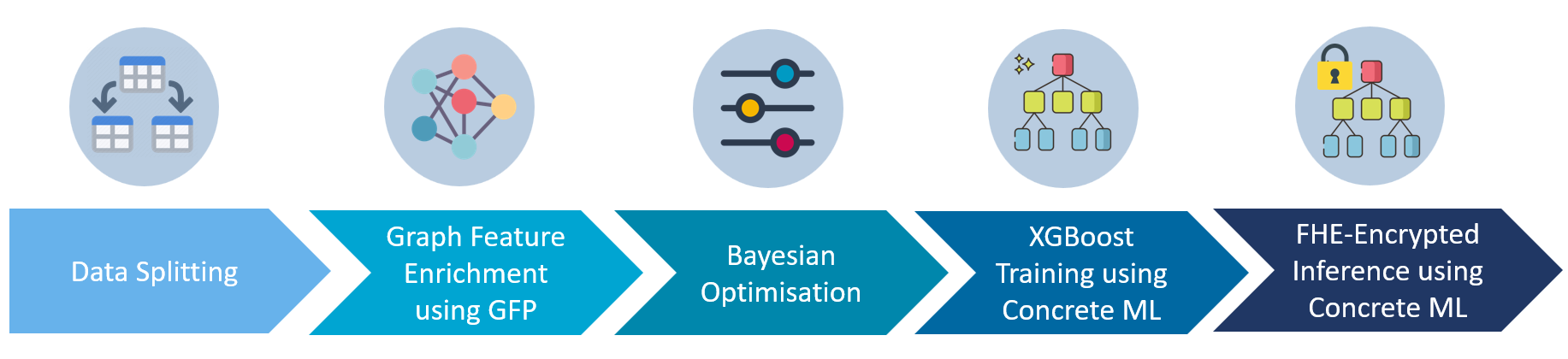}}
  \label{fig:fhe-xgb-pipeline}
  \vspace{-6pt}
  \caption{Privacy-Preserving XGBoost Pipeline}
  \vspace{-12pt}
\end{figure}

The privacy-preserving graph-based XGBoost pipeline was successfully developed. This novel approach combines the predictive power of XGBoost with the security and confidentiality guarantees provided by TFHE, while leveraging Snap ML Graph Feature Preprocessor (GFP) to enrich the model with informative graph-based features.

The XGBoost pipeline encompasses the following steps: splitting the data, enriching graph features using GFP, performing hyperparameter tuning using Bayesian optimization, training the XGBoost model and performing FHE-encrypted inference using Concrete ML.

\subsection{Data Splitting}
Data splitting was performed in a temporal manner. The transactions were ordered in ascending order of timestamps and split into train and test sets. Generally, transactions occurring before timestamp \(T_1\) are included in the training set, while transactions occurring after timestamp \(T_1\) are included in the testing set. Transactions from the same day were placed in the same set too. This method was employed to prevent data leakage and ensured that the model generalizes well to unseen data by maintaining the temporal order of transactions and avoiding information leakage from future to past data. Additionally, grouping same-day transactions enhanced the integrity of the dataset and facilitated more effective model training. The distribution ensures that the training and testing sets are representative of the overall dataset while maintaining a balanced illicit ratio for effective model training and evaluation.

\subsection{Snap ML Graph Feature Preprocessor (GFP) Setup}
To enrich the model, the Snap ML Graph Feature Preprocessor (GFP) \cite{gfp_docs_2023} was used to generate graph-based features from the dataset. Developed by Blanuša et al. \cite{blanuša2024graph}, GFP facilitates efficient and real-time feature extraction from graph-structured data. It is compatible with scikit-learn, and simplifies the creation and maintenance of in-memory graphs while extracting relevant features. To search for graph patterns, the preprocessor analyzes each edge in the input edge list to identify patterns it participates in. For every edge and pattern type, the preprocessor computes a pattern histogram, which records the count of patterns of a given size. In the AML context, these additional features can augment the predictive capability of models like XGBoost, especially in detecting suspicious transactions patterns within financial transaction graphs.

The graph-based features were extracted from the AML datasets using a time window of 86400 seconds, or 1 day. Graph-based features include fan-in, fan-out, degree-in, degree-out, scatter-gather, simple cycle and temporal cycle patterns. For the vertex-statistics-based features, the "Amount" field of the basic transaction features were used to generate the vertex-statistics-based features. Vertex-statistics-based features selected were sum, variance and skewness.

An incremental approach was adopted to assess the impact of adding various graph-based features on the performance of the XGBoost model. These features were grouped into distinct categories: (i) basic features, (ii) fan-in/fan-out features, (iii) multi-hop pattern features, encompassing scatter-gather, simple cycle, and temporal cycle patterns, (iv) vertex-statistic-based features.
 
\subsection{Hyperparameter Tuning via Bayesian Optimization}
Bayesian optimization was employed to optimize the hyperparameters of the XGBoost model. This method systematically explores the hyperparameter space by iteratively assessing the model's performance with various parameter combinations, guided by a probabilistic model of the objective function. By selecting hyperparameters for evaluation based on the model's predictions, Bayesian optimization efficiently navigates the search space and identifies the hyperparameters values that maximize the model's performance. This approach enables the identification of optimal hyperparameters with fewer evaluations compared to exhaustive grid search or random search methods.

\subsection{Training and Inference using Concrete ML}
After Bayesian optimization, the optimized XGBoost model undergoes the Concrete ML pipeline for training and FHE-encrypted inference. This process involves several key steps. First, the model is trained on plaintext, non-encrypted training data. Next, the floating-point values in the model are transformed into integers through a quantization step. The quantized model is then executed in a simulation environment to assess its accuracy under FHE and identify any necessary modifications to ensure full compatibility. Following this, the quantized model is compiled into an equivalent FHE circuit, which can be executed on encrypted data. Finally, once the appropriate cryptographic keys are generated, the quantized XGBoost model is executed on the encrypted data, enabling privacy-preserving inference using the Concrete ML framework.
\section{Experiments and Results}\label{chapt:experiments}
\subsection{Dataset}
\subsubsection{Dataset Selection.}
To train the machine learning models, a synthetic AML dataset was used to simulate money laundering scenarios, as strict privacy regulations surrounding banking data make real-world financial crime data not easily accessible. After evaluating several publicly available datasets \cite{mahootiha2020money,suzumura2021aml,altman2023realistic}, the AMLworld HI-Small dataset \cite{altman2023realistic,ibmresearch2023amldataset} was selected for its comprehensive representation of real-world money laundering activities.

The AMLworld HI-Small dataset models diverse patterns, including placement, layering, and integration, across multiple banks and currencies. Data exploration revealed 370 groups of money laundering patterns, comprising various topologies such as fan-in, fan-out, gather-scatter, cycles, and random patterns.

\subsubsection{Dataset Modification.}
Due to computational constraints posed by the slower operations inherent in Fully Homomorphic Encryption (FHE), the original AMLworld HI-Small dataset was downsized. Two modified datasets were created using undersampling and random sampling techniques respectively:
\begin{enumerate}
    \item Modified AML Dataset 1: A balanced dataset with 15,230 accounts, 10,354 transactions, and a 50\% illicit ratio.
    \item Modified AML Dataset 2: An imbalanced dataset with 9,070 accounts, 5,491 transactions, and a 5.72\% illicit ratio.
\end{enumerate}

These modified datasets provide a robust foundation for developing and evaluating machine learning pipelines to combat financial crimes, while accounting for the computational limitations of FHE.

\subsection{Environment Setup}
The experiments were conducted on a Linux system with an Intel Xeon CPU E5-1630 v4 @ 3.70GHz with 8 CPU cores, using Concrete ML version 1.4.1 to implement TFHE. As Concrete ML currently does not offer GPU support, CPU resources were utilized for the experiments.

\subsection{Experimental Overview}
Experiments were conducted on the privacy-preserving XGBoost pipeline using the 2 modified AML datasets, with inference conducted on both unencrypted (clear) and FHE-encrypted data. XGBoost served as the baseline model, and graph features were incrementally introduced in 3 main categories: (i) single-hop patterns, (ii) multi-hop patterns and (iii) vertex statistics patterns. These graph features were added onto the base dataset features as inputs for the training and testing pipelines.
\begin{table}[htbp]
\vspace{-12pt}
\centering
\caption{Pattern categories of Graph-Based Features}
\vspace{-6pt}
\renewcommand{\arraystretch}{1.2}
\begin{tabular}{|l|l|}
\hline
\textbf{Pattern Category} & \textbf{Graph-Based Features} \\
\hline
Single-Hop & Fan-in, fan-out, degree-in, degree-out \\
Multi-Hop & Scatter-gather, simple cycle, temporal cycle patterns \\
Vertex Statistics & Sum, variance, skewness of transaction amount \\
\hline
\end{tabular}
\label{tab:graph_features}
\end{table}
\vspace{-12pt}

In the experiments, the Bayesian hyperparameter tuning process iteratively samples various hyperparameter combinations across 50 iterations. A 3-fold cross-validation technique is employed in each iteration, where the dataset is divided into 3 equal folds for training and evaluation, with average performance guiding subsequent iterations of the optimization process. Once optimal hyperparameters are determined, the resulting model is integrated into the graph-based gradient boosting pipeline for FHE-based inference.

\subsection{Results and Discussion on Balanced AML Dataset 1}
The following analysis will examine the performance metrics from Table \ref{tab:performance_metrics_dataset_1} to assess the effectiveness of the Bayesian optimized XGBoost model in identifying illicit activities in the balanced AML dataset. Additionally, the discussion will explore the inference time overhead linked with FHE as depicted in Table \ref{tab:xgboost_prediction_1}.

\begin{table}[hbtp]
\vspace{-0.1cm}
\centering
\caption[Performance Metrics of Bayesian Optimized XGBoost on Dataset 1]{Performance metrics of Bayesian optimized XGBoost on Dataset 1. It highlights the effect of FHE encryption and adding graph-based features on XGBoost performance. Clear denotes inference on unencrypted data, while FHE denotes inference on FHE-encrypted data.}
\renewcommand{\arraystretch}{1.1}
\begin{tabular}{|l|c|c|c|c|c|c|c|c|}
\hline
\textbf{Input Features} & \multicolumn{2}{c|}{\textbf{Accuracy}} & \multicolumn{2}{c|}{\textbf{F1}} & \multicolumn{2}{c|}{\textbf{Precision}} & \multicolumn{2}{c|}{\textbf{Recall}} \\
\cline{2-9}
& \textbf{Clear} & \textbf{FHE} & \textbf{Clear} & \textbf{FHE} & \textbf{Clear} & \textbf{FHE} & \textbf{Clear} & \textbf{FHE} \\
\hline
\rowcolor{ForestGreen!40}
Basic features & 0.9972 & 0.9972 & 0.9978 & 0.9978 & 0.9981 & 0.9981 & 0.9975 & 0.9975 \\
\rowcolor{ForestGreen!40}
+ Single-hop & 0.9972 & 0.9972 & 0.9978 & 0.9978 & 0.9981 & 0.9981 & 0.9975 & 0.9975 \\
\rowcolor{ForestGreen!40}
+ Multi-hop & 0.9972 & 0.9972 & 0.9978 & 0.9978 & 0.9981 & 0.9981 & 0.9975 & 0.9975 \\
\rowcolor{ForestGreen!20}
+ Vertex statistics & 0.9964 & 0.9964 & 0.9972 & 0.9972 & 0.9969 & 0.9969 & 0.9975 & 0.9975 \\
\hline
\end{tabular}
\vspace{-5pt}
\label{tab:performance_metrics_dataset_1}
\end{table}

\textbf{High Performance with Basic Features.} The XGBoost model demonstrates impressive performance even with just the basic features, achieving over 99\% accuracy, F1 score, precision, and recall. This suggests that the model has effectively learned from the dataset and can make accurate predictions without the need for additional graph-based features.

\textbf{Little Effect of Graph Features.} Despite the presence of various graph patterns in the dataset, the inclusion of graph-based features did not significantly improve the model's performance. This indicates that on a balanced dataset, XGBoost may already capture the relevant features adequately without the need for additional complexity introduced by graph features.

\begin{table}[htbp]
\centering
\caption[Inference Time Metrics of Bayesian Optimized XGBoost Model on Dataset 1]{Inference time metrics of Bayesian optimized XGBoost on Dataset 1. It highlights the effect of FHE encryption and adding graph-based features on inference time. Time Ratio compares total inference time on FHE-encrypted data to unencrypted.}
\renewcommand{\arraystretch}{1.1}
\begin{tabular}{|l|c|c|c|c|c|}
\hline
 & \multicolumn{4}{c|}{\textbf{Inference Time (s)}} & \\
\cline{2-5}
\textbf{Input Features} & \multicolumn{2}{c|}{\textbf{Average (Batch)}} & \multicolumn{2}{c|}{\textbf{Total}} & \textbf{Time Ratio}\\
\cline{2-5}
& \textbf{Clear} & \textbf{FHE} & \textbf{Clear} & \textbf{FHE} & \textbf{(FHE / Clear)} \\
\hline
Basic features & 0.008414 & 1009.0963 & 0.1683 & 20181.9266 & 119926.12x \\
+ Single-hop & 0.005632 & 806.4183 & 0.1183 & 16128.3664 & 136363.27x \\
+ Multi-hop & 0.007108 & 818.8802 & 0.1422 & 16377.6037 & 115200.84x \\
+ Vertex statistics & 0.005781 & 959.1578 & 0.1156 & 19183.1552 & 165917.40x \\
\hline
\end{tabular}
\label{tab:xgboost_prediction_1}
\vspace{-10pt}
\end{table}

\textbf{FHE Encryption Overhead.} While the model's performance remained strong, the inference time for FHE-encrypted data was significantly higher than unencrypted data (Table \ref{tab:xgboost_prediction_1}), exceeding 100,000 times the unencrypted inference time. This highlights the need to explore strategies that can reduce the FHE computational overhead without compromising model performance.

Future work could focus on balancing model metrics and FHE inference time, potentially by adjusting parameters or feature sets to maintain high accuracy, precision, and recall while minimizing the encryption overhead.

\subsection{Results and Discussion on Imbalanced AML Dataset 2}

The following analysis evaluates the performance metrics (Table \ref{tab:performance_metrics_dataset_2}) and inference time (Table \ref{tab:xgboost_prediction_2}) for the imbalanced AML dataset. Here, the minority F1-score is a better measure of performance than accuracy, as the latter can be misleading by favoring the majority class. In contrast, the F1-score considers both precision and recall, offering a balanced assessment and highlighting the detection of illicit money laundering transactions.

\begin{table}[htbp]
\vspace{-2pt}
\centering
\caption[Performance Metrics of Bayesian Optimized XGBoost on Dataset 2]{Performance metrics of Bayesian optimized XGBoost on Dataset 2. It highlights the effect of FHE encryption and adding graph-based features on XGBoost performance. Clear denotes inference on unencrypted data, while FHE denotes inference on FHE-encrypted data.}
\vspace{-3pt}
\renewcommand{\arraystretch}{1.1}
\begin{tabular}{|l|c|c|c|c|c|c|c|c|}
\hline
\textbf{Input Features} & \multicolumn{2}{c|}{\textbf{Accuracy}} & \multicolumn{2}{c|}{\textbf{F1}} & \multicolumn{2}{c|}{\textbf{Precision}} & \multicolumn{2}{c|}{\textbf{Recall}} \\
\cline{2-9}
& \textbf{Clear} & \textbf{FHE} & \textbf{Clear} & \textbf{FHE} & \textbf{Clear} & \textbf{FHE} & \textbf{Clear} & \textbf{FHE} \\
\hline
\rowcolor{ForestGreen!20}
Basic features & 0.9016 & 0.9016 & 0.3056 & 0.3056 & 0.7857 & 0.7857 & 0.1897 & 0.1897 \\
\rowcolor{ForestGreen!60}
+ Single-hop & 0.9094 & 0.9094 & 0.3867 & 0.3867 & 0.8529 & 0.8529 & 0.2500 & 0.2500 \\
\rowcolor{ForestGreen!45}
+ Multi-hop & 0.9075 & 0.9075 & 0.3733 & 0.3733 & 0.8235 & 0.8235 & 0.2414 & 0.2414  \\
\rowcolor{ForestGreen!30}
+ Vertex statistics & 0.9035 & 0.9035 & 0.3467 & 0.3467 & 0.7647 & 0.7647 & 0.2241 & 0.2241 \\
\hline
\end{tabular}
\label{tab:performance_metrics_dataset_2}
\vspace{-10pt}
\end{table}

\textbf{Lower Performance on Imbalanced Dataset.} For the imbalanced dataset, the model faced challenges in accurately identifying illicit activities, evident from the drop in F1-score and recall compared to the balanced dataset. The imbalanced nature negatively impacted the F1-score, a key metric for such scenarios.

\textbf{Improved Performance with Graph-Based Features.} Despite the challenges posed by the imbalanced dataset, incorporating graph-based features improved the model's predictive capabilities. The addition of single-hop features led to the highest accuracy, F1-score, precision, and recall among all feature sets, outperforming the basic features. However, adding more advanced multi-hop and vertex statistics features resulted in a slight performance decrease compared to the single-hop features, though they still outperformed the basic features. Further investigation is required to understand the reasons behind this performance change when incorporating the additional graph-based feature categories.

\textbf{Consistent Performance in Encrypted Inference.} Moreover, the consistency in performance between unencrypted and FHE-encrypted inference highlights the robustness of Concrete-TFHE in maintaining model accuracy, F1-score, precision and recall while preserving privacy. This underscores the potential of Concrete ML in preserving privacy without sacrificing predictive performance, a crucial aspect in sensitive domains like financial transactions.

\begin{table}[htbp]
\vspace{-6pt}
\centering
\caption[Inference Time Metrics of Bayesian Optimized XGBoost Model on Dataset 2]{Inference time metrics of Bayesian optimized XGBoost on Dataset 2. It highlights the effect of FHE encryption and adding graph-based features on inference time. Time Ratio compares total inference time on FHE-encrypted data to unencrypted.}
\vspace{-3pt}
\renewcommand{\arraystretch}{1.1}
\begin{tabular}{|l|c|c|c|c|c|}
\hline
 & \multicolumn{4}{c|}{\textbf{Inference Time (s)}} & \\
\cline{2-5}
\textbf{Input Features} & \multicolumn{2}{c|}{\textbf{Average (Batch)}} & \multicolumn{2}{c|}{\textbf{Total}} & \textbf{Time Ratio}\\
\cline{2-5}
& \textbf{Clear} & \textbf{FHE} & \textbf{Clear} & \textbf{FHE} & \textbf{(FHE / Clear)} \\
\hline
Basic features & 0.003041 & 270.1013 & 0.02433 & 2160.8107 & 88822.54x \\
+ Single-hop & 0.003055 & 64.9026 & 0.02444 & 519.2208 & 21246.95x \\
+ Multi-hop & 0.002270 & 34.1322 & 0.01816 & 273.0572 & 15038.47x\ \\
+ Vertex statistics & 0.003607 & 121.0173 & 0.02886 & 968.1386 & 33546.76x \\
\hline
\end{tabular}
\label{tab:xgboost_prediction_2}
\vspace{-10pt}
\end{table}

\textbf{Effect of Feature Complexity on Inference Time.}
Interestingly, as graph-based feature complexity increased, inference time decreased for both clear and FHE-encrypted data, suggesting that the overhead of advanced features may be offset by their ability to streamline the inference process, resulting in more efficient model predictions. Further analysis is needed to validate this observation. Other potential factors contributing to this trend could include the optimization of feature representations, the inherent parallelism in graph-based computations, or the effectiveness of the Bayesian optimization process. 

Overall, the integration of graph-based features showed promise in enhancing the model's ability to detect illicit activities. Future research could focus on optimizing model performance for imbalanced scenarios.

\section{Conclusion}
This paper proposed a novel privacy-preserving approach for collaborative AML detection using FHE. Two key pipelines were developed:
\begin{enumerate}
    \item A privacy-preserving GNN pipeline that explored the integration of GIN with TFHE, where optimization techniques like quantization and pruning were used in attempts to render the GNN FHE-compatible.
    \item A privacy-preserving graph-based XGBoost pipeline that leveraged the Graph Feature Preprocessor (GFP) to enhance the model's predictive performance on the AML datasets. Experiments demonstrated the XGBoost model's ability to achieve over 99\% accuracy, F1-score, precision, and recall on the balanced dataset, with the incorporation of graph-based features improving the F1-score by 8\% on the imbalanced dataset.
\end{enumerate}

A key strength of the proposed approach was the consistent performance of the XGBoost model in both unencrypted and FHE-encrypted inference settings, highlighting the robustness of the Concrete-TFHE framework in preserving privacy without compromising predictive capabilities. However, it also underscored the need to balance the trade-off between privacy preservation and computational efficiency, as FHE-encrypted inference incurred significant overhead.

\section{Future Work}
This work lays the foundation for innovative privacy-preserving approaches to AML detection. Future research could focus on:
\begin{enumerate}
    \item Improving FHE compatibility and performance of models.
    \item Investigating alternative privacy-preserving methods, such as differential privacy or multi-party computation, to enhance privacy-preserving capabilities.
    \item Assessing scalability of solutions to accomodate larger datasets.
\end{enumerate}

By addressing these future directions, the privacy-preserving machine learning approaches developed in this paper can be further advanced, contributing to the ongoing efforts to safeguard financial systems against illicit activities while preserving data privacy. Additionally, the insights and techniques explored in this paper may have the potential for broader applications in other domains that require privacy-preserving collaborative machine learning.


\begin{credits}
\subsubsection{\ackname} The authors would like to thank Alka Luqman for her contributions during the debugging of the GNN pipeline.

\subsubsection{\discintname}
The authors have no competing interests to declare that are relevant to the content of this article.
\end{credits}
%
%
%
%



\appendix
\section{Key Considerations in Collaborative FHE Architecture}
This appendix outlines some of the key considerations in the design of the collaborative FHE architecture.

\subsection{Advantages of Collaborative FHE over MPC}
Below details the key advantages of our proposed collaborative FHE architecture in comparison to traditional Multi-Party Computation (MPC) methods in the context of AML applications.

In MPC, data fragmentation is an important requirement where multiple parties must divide and share their data to perform computations on distributed fragments. This process necessitates continuous coordination and communication among all participating FIs, resulting in significant overhead and complexity. Each party must remain online and actively participate in every computation step, making the process highly interactive and less scalable.

In contrast, our collaborative FHE architecture does not rely on data fragmentation and allows for non-interactive processing of encrypted data. Once an FI encrypts its data using a public key, computations can proceed independently without further interaction. This non-interactive approach not only simplifies data handling but also enhances efficiency by allowing FIs to contribute data without needing to be online throughout the entire computation process.

Moreover, our FHE architecture enables multiple FIs to encrypt and submit their data for secure computation while requiring only the inquiry FI to decrypt the final result. This significantly reduces communication overhead, resulting in improved scalability compared to MPC, which incurs considerable overhead due to constant data exchanges among participants.

\subsection{Public Key Sharing}
Our FHE architecture effectively manages public key sharing and minimizes associated network overhead.

\begin{itemize}
    \item \textbf{Initial Key Sharing:} The inquiry FI's public key can be securely distributed at the start of a session or when a new FI initiates an inquiry. Once the key is shared, multiple transactions can be encrypted under the same key, reducing the need for repeated re-encryption for individual inquiries. This streamlines the processing of multiple data transactions within a single inquiry session.
    
    \item \textbf{Managing Changes in Inquiry FI:} If a new inquiry FI is introduced, its public key is shared efficiently with participating FIs, requiring minimal overhead as the key exchange only involves secure transmission of a small piece of data.
    
    \item \textbf{Session-Based Key Exchange:}  The architecture supports session-based key exchanges, enabling each inquiry FI to generate a new session key or update its public key for each inquiry. This approach ensures that only new transactions require encryption with the new key, while existing encrypted data can continue to be processed without interruption.
\end{itemize}

\section{Detailed characteristics of modified AML datasets}
This appendix provides additional details on the modifications made to the original AMLworld HI-Small dataset.

\subsection{Modified AML Dataset 1}
Undersampling was used to create a balanced dataset of 10,354 transactions and 15,230 accounts, preserving rows associated with the 370 money laundering pattern groups in the original dataset (Table \ref{table:pattern_groups}). This reduced the dataset size for FHE computations while balancing the 50
\begin{table}[htbp]
    \vspace{-12pt}
    \centering
    \caption{Money Laundering Pattern Groups in AMLworld HI-Small dataset}
    \label{table:pattern_groups}
    \begin{tabular}{|l|r|}
    \hline
    Pattern & Count \\
    \hline
    Fan-In & 61 \\
    Fan-Out & 80 \\
    Gather-Scatter & 77 \\
    Cycle & 82 \\
    Random & 70 \\
    \hline
    Total Patterns & 370 \\
    \hline
    \end{tabular}
    \vspace{-12pt}
\end{table}

\begin{table}[htbp]
\vspace{-12pt}
    \centering
    \caption{Characteristics of Modified AML Dataset 1}
    \renewcommand{\arraystretch}{1.2}
    \begin{tabular}{|l|c|}
        \hline
        \textbf{Characteristics} & \textbf{Value} \\
        \hline
        Number of accounts & 15,230 \\
        Number of transactions & 10,354 \\
        Illicit money laundering ratio & 50\% \\
        \hline
    \end{tabular}
    \label{tab:modified_dataset}
    \vspace{-12pt}
\end{table}

\subsection{Modified AML Dataset 2}
A smaller, highly imbalanced dataset was created by randomly sampling 40 out of the 370 pattern groups to represent illicit transactions (Table \ref{table:pattern_groups_2}). This resulted in 5,491 transactions across 9,070 accounts, with a 5.72\% illicit ratio (Table \ref{tab:modified_dataset_2}).
\begin{table}[htbp]
    \vspace{-12pt}
    \centering
    \caption{Money Laundering Pattern Groups in Modified AML Dataset 2}
    \begin{tabular}{|l|r|}
    \hline
    Pattern & Count \\
    \hline
    Fan-In & 8 \\
    Fan-Out & 7 \\
    Gather-Scatter & 9 \\
    Cycle & 8 \\
    Random & 8 \\
    \hline
    Total Patterns & 40 \\
    \hline
    \end{tabular}
    \label{table:pattern_groups_2}
    \vspace{-12pt}
\end{table}

\begin{table}[hbtp]
\vspace{-2pt}
    \centering
    \caption{Characteristics of Modified AML Dataset 2}
    \renewcommand{\arraystretch}{1.2}
    \begin{tabular}{|l|c|}
        \hline
        \textbf{Characteristics} & \textbf{Value} \\
        \hline
        Number of accounts & 9070 \\
        Number of transactions & 5491 \\
        Illicit money laundering ratio & 5.72\% \\
        \hline
    \end{tabular}
    \label{tab:modified_dataset_2}
    \vspace{-4pt}
\end{table}

\section{Train-Test Split Distribution}
The datasets were split temporally, with the training set comprising 75-81\% of the data (Tables \ref{tab:train_test_split_1} and \ref{tab:train_test_split_2}).
\begin{table}[htbp]
    \vspace{-12pt}
    \centering
    \caption{Train-Test Split Distribution for Modified AML Dataset 1}
    \renewcommand{\arraystretch}{1.2}
        \begin{tabular}{|lcc|}
        \hline
        & \textbf{Percentage of Dataset} & \textbf{Illicit Ratio} \\
        \hline
        Train Sample & 75.85\% & 45.47\% \\
        Test Sample & 24.15\% & 64.16\% \\
        \hline
    \end{tabular}
    \label{tab:train_test_split_1}
    \vspace{-12pt}
\end{table}

\begin{table}[htbp]
    \vspace{-12pt}
    \centering
    \caption{Train-Test Split Distribution for Modified AML Dataset 2}
    \renewcommand{\arraystretch}{1.2}
        \begin{tabular}{|lcc|}
        \hline
        & \textbf{Percentage of Dataset} & \textbf{Illicit Ratio} \\
        \hline
        Train Sample & 81.50\% & 4.42\% \\
        Test Sample & 18.50\% & 11.42\% \\
        \hline
    \end{tabular}
    \label{tab:train_test_split_2}
    \vspace{-4pt}
\end{table}

\section{Definition of Graph-Based Features}
Below are an elaboration of the graph-based features.
\begin{itemize}
    \item \textbf{Fan-in/fan-out} reveals the inbound and outbound connectivity of vertices, aiding in understanding recipient and initiator vertices within the graph.

    \item \textbf{Degree-in/degree-out} quantifies the total inbound and outbound edges for each vertex, providing a holistic view of vertex connectivity.

    \item \textbf{Scatter-gather patterns} detects dispersion and aggregation behaviors across the graph, offering insights into information propagation and aggregation.

    \item \textbf{Simple cycles} identifies closed paths within the graph, highlighting recurring patterns or loops indicative of repetitive processes.

    \item \textbf{Temporal cycles} uncovers cyclic patterns with temporal dimensions, offering insights into recurring activities over time.
\end{itemize}

Below are an elaboration of the vertex-statistics-based features.
\begin{itemize}
    \item \textbf{Sum} provides insight into the overall volume of financial activity. High sums may indicate potentially suspicious behavior, such as large-scale transactions or money laundering schemes.
    \item \textbf{Variance} reflects the dispersion or spread of transaction amounts around the mean. High variance may suggest irregular or unpredictable patterns in financial transactions, indicative of fraudulent activities.
    \item \textbf{Skewness} measures the asymmetry of the distribution of transaction amounts. Positive skewness suggests a longer tail towards higher values, while negative skewness suggests a longer tail towards lower values. Extreme positive skewness may indicate unusual transaction patterns that require further investigation.
\end{itemize}

\section{Hyperparameter Tuning Seach Space}
The Bayesian optimization search space for XGBoost hyperparameters is detailed in Table \ref{tab:param_ranges}.
\begin{table}[htbp]
\vspace{-12pt}
    \centering
    \caption{Search Space for Bayesian Optimization of XGBoost Hyperparameters}
    \renewcommand{\arraystretch}{1.2}
    \begin{tabular}{|l|c|}
        \hline
        \textbf{Parameter} & \textbf{Search Space Range} \\
        \hline
        \texttt{n\_estimators} & (5, 30) \\
        \texttt{max\_depth} & (2, 12) \\
        \texttt{learning\_rate} & (0.003, 0.1) \\
        \texttt{colsample\_bytree} & (0.5, 1) \\
        \hline
    \end{tabular}
    \label{tab:param_ranges}
    \vspace{-4pt}
\end{table}

\clearpage
\section{Additional Privacy-Preserving XGBoost Experiments}
Experiments varying n\_estimators, max\_depth, and n\_bits provided insights into balancing model performance and FHE inference time (Tables \ref{tab:nestimators}, \ref{tab:maxdepth}, \ref{tab:nbits}).

\begin{table}[htbp]
\vspace{-10pt}
\centering
\caption[Performance and Time Metrics of XGBoost Models with Varying n\_estimators.]{Performance and time metrics of XGBoost models with varying n\_estimators.  Experiment was conducted on Dataset 1 at fixed n\_bits = 3, learning\_rate = 0.07, colsample\_bytree = 0.98, max\_depth = 3.}
\renewcommand{\arraystretch}{1.1}
\begin{tabular}{|c|c|c|c|c|c|c|c|}
\hline
\textbf{n\_estimator} & \multicolumn{2}{c|}{\textbf{Accuracy}} & \multicolumn{2}{c|}{\textbf{F1-Score}} & \multicolumn{2}{c|}{\textbf{Inference Time (s)}}& \textbf{Time} \\
\cline{2-7} 
& \textbf{Clear} & \textbf{FHE} & \textbf{Clear} & \textbf{FHE} & \textbf{Clear} & \textbf{FHE} & \textbf{Ratio}\\
\hline
5 & 0.6428 & 0.6428 & 0.7822 & 0.7822 & 0.009405 & 1401.32 & 149000x \\
10 & 0.6428 & 0.6428 & 0.7822 & 0.7822 & 0.013848 & 2222.64 & 160504x \\
50 & 0.6432 & 0.6432 & 0.7824 & 0.7824 & 0.042257 & 26595.60 & 629379x \\
100 & 0.6483 & 0.6483 & 0.7846 & 0.7846 & 0.205946 & 55523.89 & 269604x \\
200 & 0.6483 & 0.6468 & 0.7846 & 0.7833 & 0.174605 & 65481.47 & 375027x \\
\hline
\end{tabular}
\label{tab:nestimators}
\end{table}

\begin{table}[htbp]
\vspace{-12pt}
\centering
\caption[Performance and Time Metrics of XGBoost Models with Varying max\_depth.]{Performance and time metrics of XGBoost models with varying max\_depth.  Experiment was conducted on Dataset 1 at fixed n\_bits = 3, learning\_rate = 0.07, colsample\_bytree = 0.98, n\_estimator = 20.}
\renewcommand{\arraystretch}{1.1}
\begin{tabular}{|c|c|c|c|c|c|c|c|}
\hline
\textbf{max\_depth} & \multicolumn{2}{c|}{\textbf{Accuracy}} & \multicolumn{2}{c|}{\textbf{F1-Score}} & \multicolumn{2}{c|}{\textbf{Inference Time (s)}}& \textbf{Time Ratio} \\
\cline{2-7}
& \textbf{Clear} & \textbf{FHE} & \textbf{Clear} & \textbf{FHE} & \textbf{Clear} & \textbf{FHE} & \\
\hline
1 & 0.3584 & 0.3584 & 0.0000 & 0.0000 & 0.019856 & 4014.30 & 202173.40x \\
4 & 0.5540 & 0.5540 & 0.6809 & 0.6809 & 0.121287 & 29193.94 & 240701.55x \\
7 & 0.5540 & 0.5540 & 0.6809 & 0.6809 & 1.873080 & 47420.48 & 25316.84x \\
10 & 0.5540 & 0.5540 & 0.6809 & 0.6809 & 0.228830 & 40407.25 & 176581.86x \\
13 & 0.5540 & 0.5540 & 0.6809 & 0.6809 & 0.184337 & 41289.46 & 223988.81x \\
\hline
\end{tabular}
\label{tab:maxdepth}
\end{table}

\begin{table}[htbp]
\vspace{-12pt}
\centering
\caption[Performance and Time Metrics of XGBoost Models with Varying n\_bits.]{Performance and time metrics of XGBoost models with varying n\_bits. Experiment was conducted on Dataset 1 at fixed learning\_rate = 0.07, colsample\_bytree = 0.98, n\_estimator = 20, max\_depth = 3.}
\renewcommand{\arraystretch}{1.1}
\begin{tabular}{|c|c|c|c|c|c|c|c|}
\hline
\textbf{n\_bits} & \multicolumn{2}{c|}{\textbf{Accuracy}} & \multicolumn{2}{c|}{\textbf{F1-Score}} & \multicolumn{2}{c|}{\textbf{Inference Time (s)}} & \textbf{Time Ratio} \\
\cline{2-7}
& \textbf{Clear} & \textbf{FHE} & \textbf{Clear} & \textbf{FHE} & \textbf{Clear} & \textbf{FHE} & \\
\hline
2 & 0.4540 & 0.4540 & 0.4553 & 0.4553 & 0.042825 & 7244.70 & 169168.00x \\
3 & 0.5540 & 0.5540 & 0.6809 & 0.6809 & 0.055623 & 11766.27 & 211534.87x \\
4 & 0.6432 & 0.6432 & 0.7824 & 0.7824 & 0.055750 & 30344.95 & 544307.47x \\
8 & 0.6428 & 0.6428 & 0.7821 & 0.7821 & 0.023308 & 32139.84 & 1378944.61x \\
\hline
\end{tabular}
\label{tab:nbits}
\vspace{-12pt}
\end{table}

\end{document}